\documentclass[10pt,twocolumn]{article}

\usepackage[utf8]{inputenc}
\usepackage[T1]{fontenc}

\usepackage{graphicx}
\usepackage{pdfpages}

\usepackage{lipsum}
\usepackage{xcolor}

\usepackage{authblk}
\usepackage[margin=3cm]{geometry}
\usepackage{mathtools, cuted}

\usepackage{bm}
\usepackage{amsfonts}
\usepackage{amsmath}
\usepackage{amssymb}
\usepackage{mathrsfs} 
\usepackage{siunitx}

\title{Dynamics of bubbles spontaneously entering in a tube}
\author[1]{Alexis Commereuc}
\author[1]{Manon Marchand}
\author[1]{Emmanuelle Rio}
\author[1]{François Boulogne}
\affil[1]{Université Paris-Saclay, CNRS, Laboratoire de Physique des Solides, 91405, Orsay, France.}

\date{\today}

\begin{document}

\twocolumn[
    \begin{@twocolumnfalse}
        \maketitle
        \begin{abstract}
            When an open tube of small diameter touches a bubble of a larger diameter, the bubble spontaneously shrinks and pushes a soap film in the tube.
            We characterize the dynamics for different bubble sizes and number of soap films in the tube.
We rationalize this observation from a mechanical force balance involving the Laplace pressure of the bubble and the viscous force from the advancing soap lamellae in the tube.
We propose a numerical resolution of this model, and an analytical solution in an asymptotic regime.
These predictions are then compared to the experiments.
The emptying duration is primarily affected by the initial bubble to tube diameter ratio and by the number of soap films in the tube.
        \end{abstract}
    \end{@twocolumnfalse}
]

%
%
\section{Introduction}

Spontaneous entrance of a liquid in capillary tubes is a well studied problem.
The equilibrium solution of capillary rise is the well known Jurin's law and was derived in the 18th century \cite{Jurin1719}.
It is then two centuries later that the pioneering work by Bell, Lucas and Washburn allowed the understanding of the dynamics of this problem.
They considered the viscous dissipation in the liquid flow driven by the capillary suction \cite{Bell1906,Lucas1918,Washburn1921}.
More recently, this problem has been revisited to provide a finer description of the role of inertia and  the possibility of interfacial oscillations for fluids of extremely low viscosity  was revealed \cite{Quere1997a,Quere1999a}.

Liquid invading capillary tubes has been extended also to drops.
Wang \textit{et al.} performed numerical simulations on the penetration of drop in non-wetting capillaries \cite{Wang2017} that has been studied experimentally by
Willmott \textit{et al.} highlighting the effect of the wetting conditions and drop size to tube ratio \cite{Willmott2011}.
For drops, the dynamics is dominated by inertia and capillarity, with a noticeable correction due to viscous forces \cite{Willmott2020}.

Besides drops, bubbles are also prone to shrinkage due to the Laplace pressure for which
Leidenfrost wrote observations in a monograph entitled \textit{De aquae communis nonnullis qualitatibus tractatus} published in 1756 \cite{Leidenfrost1756}.
Leidenfrost is renowned for describing a phenomenon where a liquid drop is placed on a hot surface whose temperature is larger than the boiling temperature of the liquid
This \textit{Treatise on the Properties of Common Water} not only contains a text on levitating drops, but also a less known part on soap films and bubbles.
Leidenfrost commented on the soap film thickness, colored fringes, and also on surface tension and induced dynamics.
Notably, Leidenfrost reports observations in the paragraph 46 on the bubble shrinkage when placed at the extremity of a tube.

Another well known configuration where bubble shrinkage is encountered is foam aging, in which the smaller bubbles tend to empty into the larger ones because of their higher Laplace pressure.
In such a situation, the motor, which is the pressure difference between the bubbles is balanced by the gas diffusion in the liquid film \cite{Cantat2013b}.

Another situation where a bubble shrinkage dynamics is observed has been reported recently by Clerget \textit{et al.} who  studied experimentally and theoretically the emptying of a hemispherical bubble through a pierced surface \cite{Clerget2021}.
The air motion is driven by the Laplace pressure in the bubble limited by the air flow through the orifice and eventually by the friction of the bubble on the surface.

 \begin{figure}[tp]
     \centering
     \includegraphics[width=.8\linewidth]{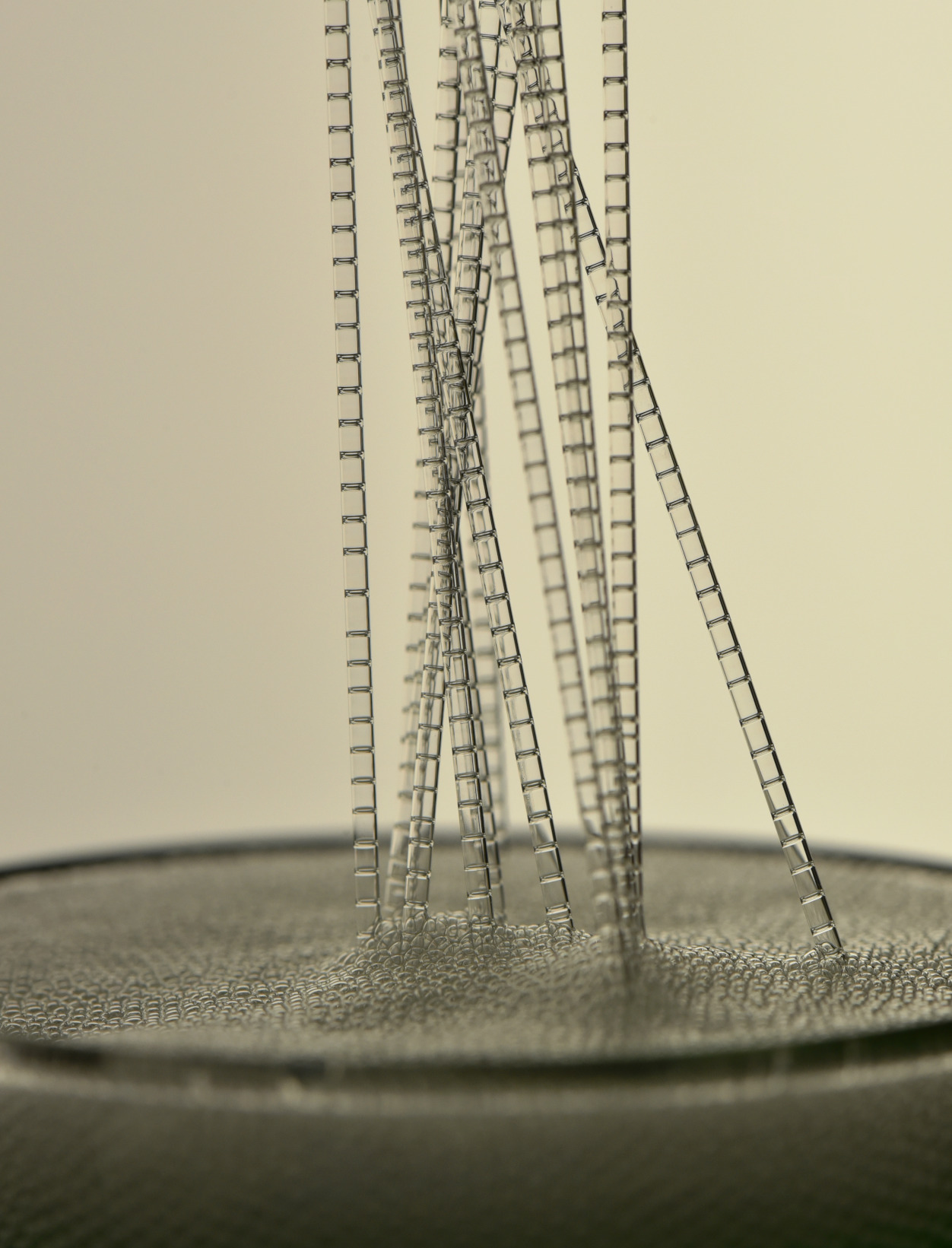}
     \caption{
     Bamboo foam in capillary tubes placed into contact of the surface of a glass containing a monodisperse foam.
     The tube diameter is 1 mm.
     }
     \label{fig:bamboo_foam_illustration}
 \end{figure}

In this article, we propose to investigate the situation closely related to the observations made by Leidenfrost: the shrinkage of a bubble placed at the end of a tube of a smaller diameter.
A careful observation reveals that the entering motion shows a characteristic pattern of acceleration-deceleration, as the bubble shrinks.
This scenario can be repeated by bringing additional bubbles to the tube extremity or even more simply by dipping the tube in a foam.
The successive bubble draining leads to the formation of a so-called bamboo foam as shown in figure \ref{fig:bamboo_foam_illustration}.

These observations are quantified and rationalized through experimental measurements in which bubbles are manually put in contact with the tube, described in Section~\ref{sec:exp}.
A simple model for the dynamics of the soap lamellae formed in the tube is presented in Section~\ref{sec:model}.
This model is solved as is numerically, and with some approximations analytically.
We show that the problem differs from the one of the spontaneous entrance of a drop in a capillary tube \cite{Willmott2011,Willmott2020} because first, inertia is negligible for bubbles and second, viscous dissipation is due to the motion of the lamellae.

%
%
%

\section{Experimental protocol and qualitative observations}\label{sec:exp}

We prepare a soap solution by diluting a commercial dish washing soap (Fairy purchased at a concentration in anionic surfactant of 5–15 \%) at a concentration of 10 wt.\% in pure water.
The liquid-vapor surface tension is $\gamma = 24.5 \pm 0.1$ mN/m and the viscosity is $\mu_{\ell} = 1.0 \pm 0.2~ $mPa$\cdot$s.

\begin{figure*}
    \centering
    \includegraphics[width=0.8\textwidth]{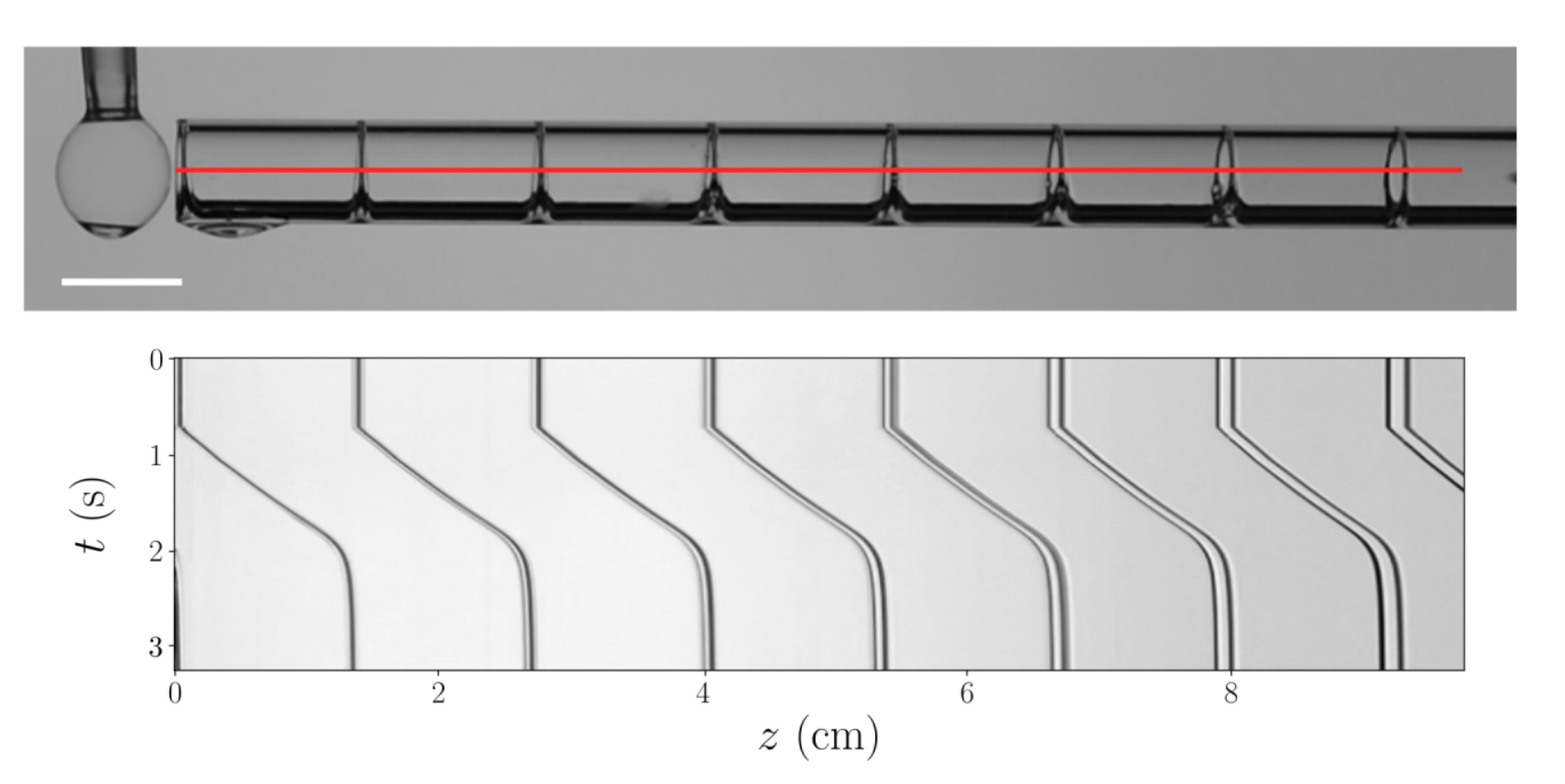}
    \caption{
    Bubble approached at the extremity of a tube where a bamboo foam is already made from previous bubbles.
    The plot is a space-time diagram along the red line shown on the photograph that illustrates the dynamics of soap films inside the tube when a bubble empties.
    The constant distance between soap films is an experimental evidence of the incompressible air flow.
    The scale bar represents 1~cm.
    See video in the Supplementary Information.
    }
    \label{fig:spacetime}
\end{figure*}

A glass tube (Jeulin) of inner radius $a = 3.1$ mm and length $L=$ 0.75 m is pre-wetted with the soap solution and held horizontally.
Then, a bubble produced with a syringe of controlled initial volume $\Omega_{\rm bubble}$ is placed at one extremity of the tube as shown in figure~\ref{fig:spacetime}.
We consider the situation of an initial radius $R_0$, associated to an initial volume $\Omega_{\rm bubble}=\frac{4}{3} \pi R_0^3$, larger than the tube radius $a$.
The bubble empties in the tube such that a soap lamella enters in the tube.
This lamella stops when the bubble is fully entered in the tube.
A movie from which this time series has been extracted is available in Supplementary Materials.

When the motion is completed, a new bubble of the same volume is placed following the same procedure.
A new lamella is produced and the one originally present in the tube is pushed forward.
The Mach number, defined as ratio of the lamella velocity and the speed of sound in air, is between $10^{-6}$ and $10^{-5}$, which is much smaller than unity.
Thus, the air flow is at a very good approximation incompressible such that both lamellae move at the same velocity and the distance between them is set by conservation of the initial bubble air volume as shown in the space-time diagram in figure~\ref{fig:spacetime}.

\begin{figure}[ht]
     \centering
     \includegraphics[width=.99\linewidth]{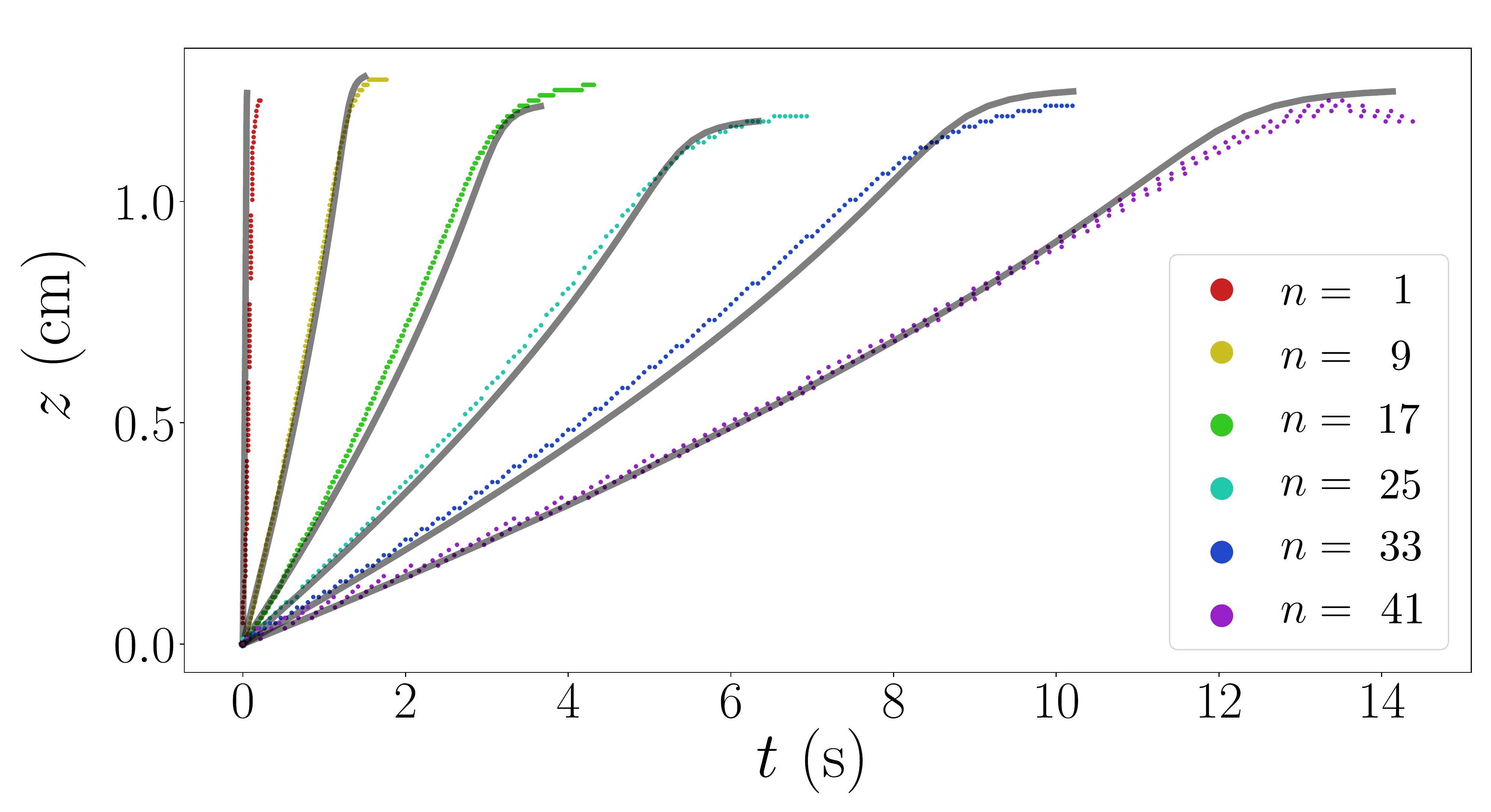}\\
     \caption{Dynamics for different number $n$ of soap lamellae in the tube for bubbles of initial volume $\Omega_{\rm bubble} = 0.42 \pm 0.03$~mL.
     Experimental measurements are represented by colored dots and the numerical solution are the black solid lines for $\xi=32$.
     }
     \label{fig:bubble_dynamics_example}
\end{figure}

The addition of bubbles is repeated until most of the tube is filled with soap lamellae.
The dynamics is recorded with a DSLR camera (Nikon D750 with a 50~mm objective) at 60 fps.
We denote $z$ the position of the nearest soap film from the tube entrance.
We measure the time evolution of this position as a function of the number $n$ of soap films in the tube.
In figure~\ref{fig:bubble_dynamics_example}, we plot the experimental position $z(t)$ where color correspond to different $n$ values and for $R_0/a \simeq 1.5$.

 By design of the experiment, all the bubbles have the same volume.
 Consequently, the same final $z$ positions are reached once the bubbles are totally inside the tube.
 \textit{A contrario}, the time taken to reach this final position increases as the number of preceding bubbles -- and thus of soap lamellae contributing to friction -- increases.
We observe two regimes: the soap film moves with a slight acceleration in the first regime and decelerates in a second regime to reach the final position.

%
%
%

\section{Model}\label{sec:model}

We propose to rationalize these observations with a simple model.
Then, the predictions of this model will be compared quantitatively to the experiments.

\subsection{Volume conservation}

 \begin{figure}
     \centering
     \includegraphics[width=.90\linewidth]{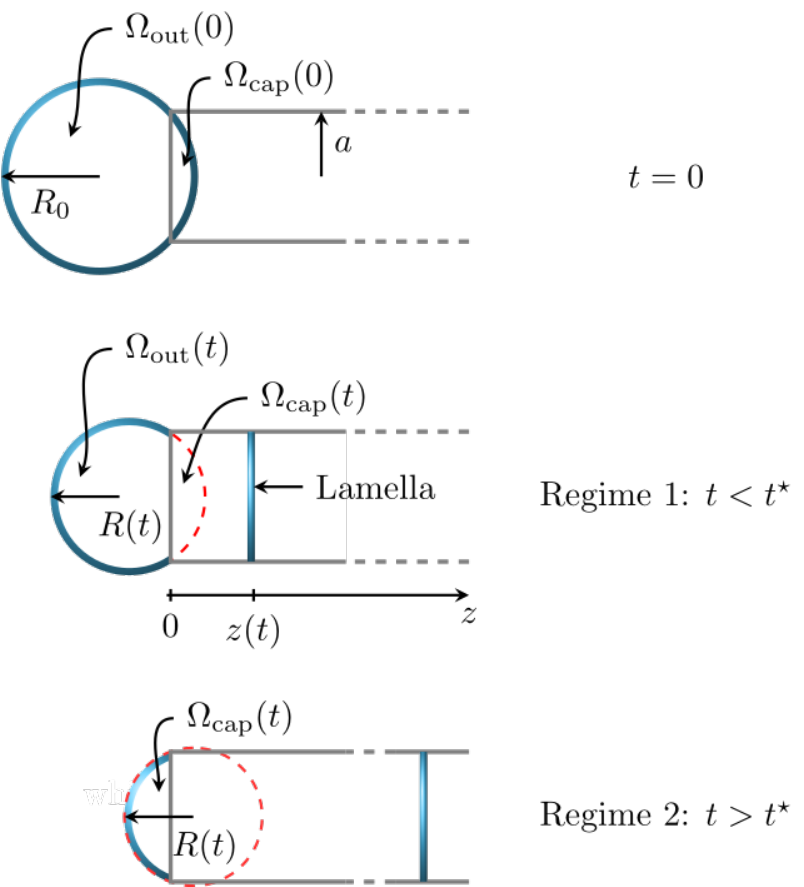}
     \caption{
     Schematics of the vanishing bubble in a tube at $t=0$, in Regime 1, and in Regime 2.
     }
     \label{fig:model_schematic}
 \end{figure}

During the emptying dynamics, the bubble shape is modeled at any time as a portion of sphere, as depicted in figure~\ref{fig:model_schematic}, because this shape results from the minimization of surface tension energy.
Initially, the bubble volume can be decomposed as $ \Omega_{\rm bubble} = \frac{4}{3} \pi R_0^3 = \Omega_{\rm out}(t=0) + \Omega_{\rm cap}(t=0) $ where $\Omega_{\rm out}$ is the volume outside the tube and $\Omega_{\rm cap}$ is the complementary part in the tube.
The volume of the cap is defined as
\begin{equation}
\Omega_{\rm cap}(t) = \frac{\pi R(t)^3}{3} \left(1 - {\cal G}\right)^2 \left(2 + {\cal G}\right),
\end{equation}
with ${\cal G}(a, R(t))=\sqrt{1 - a^2/R(t)^2}$.

From geometrical considerations (Fig.~\ref{fig:model_schematic}), the time evolution of the volume $\Omega_{\rm out}(t)$ outside the tube follows two regimes \cite{Jackson2010}.
The bubble radius of curvature first decreases until $R=a$ during a first regime, and then increases toward a zero curvature in a second regime.
We denote $t^\star$ the switching time between Regimes 1 and 2 that corresponds to $R(t^\star)=a$.
In Regime 1, we have
\begin{equation}
 \Omega^{\rm Regime1}_{\rm out}(t) = \frac{4}{3} \pi R(t)^3 - \Omega_{\rm cap}(t),\label{eq:Volume_reg1}
\end{equation}
and thereafter, in Regime 2,
\begin{equation}
     \Omega^{\rm Regime2}_{\rm out}(t) = \Omega_{\rm cap}(t).\label{eq:Volume_reg2}
\end{equation}

Since the Laplace pressure in the bubble is negligible compared to the atmospheric pressure, the expansion of air due to the pressure variation is also negligible.
Thus, the volume conservation satisfies

\begin{equation}\label{eq:volume_conservation}
    \Omega_{\rm out}(t) + z(t) \pi a^2 = \Omega_{\rm out} (0),
\end{equation}
where $z(t)$ is the position of the soap film with an origin at the entrance of the tube.
The initial condition is $z(0) = 0$.
Equation \ref{eq:volume_conservation} provides a relation between the bubble radius $R(t)$ and the lamella position $z(t)$.

\subsection{Force balance}

Due to the bubble curvature, the pressure difference between the enclosed air and the atmospheric pressure is given by the Laplace pressure
\begin{equation}\label{eq:laplace_pressure}
    P_{\rm cap} = \frac{4\gamma}{R(t)},
\end{equation}
where $R(t)$ is the bubble radius.
This Laplace pressure pushes one lamella in the tube or several lamellae, if some were initially present (see figure~\ref{fig:spacetime}).

The motion of a lamella is associated to a viscous stress dissipation that writes \cite{Bretherton1961,Hirasaki1985,Kornev1999,Raufaste2009,Cantat2013a}
\begin{equation}\label{eq:SINGLEBUBBLE_bretherton}
    \sigma_{\rm film} = 2 \xi \frac{\gamma}{a} \left(  \frac{\mu_\ell \dot z}{\gamma}  \right)^{2/3},
\end{equation}
where $\xi$ is a numerical prefactor varying quite significantly with the geometry, the quantity of liquid and the physical-chemistry of the soap films \cite{Raufaste2009,Cantat2013a}.

Therefore, the stress balance between the Laplace pressure and the opposing dissipation writes $P_{\rm cap} = n\,\sigma_{\rm film}$ where $n\geq 1$ is the number of lamellae.

We define two characteristic parameters: the traveling distance $z_{\rm r} = \Omega_{\rm bubble} / \pi a^2 $ and  the timescale $\tau =  \mu_\ell z_{\rm r} \xi^{3/2} / \gamma$.
Then, the stress balance writes in a dimensionless form
\begin{equation}\label{eq:adim_stress_balance}
   \frac{n}{2}    \tilde{R} \tilde{\dot z}^{2/3} = 1,
\end{equation}
where $\tilde R = R/a$ and $\tilde{\dot z} = \dot{z} \, \tau / z_{\rm r}$.

The volume conservation defined in equation \ref{eq:volume_conservation} becomes
\begin{equation} \label{eq:adim_volume_conservation}
        \tilde{z} = 1 - \frac{\tilde \Omega_{\rm out}(\tilde t)}{\tilde \Omega_{\rm out}(0)},
\end{equation}
with $\tilde \Omega_{\rm out} = \Omega_{\rm out} / a^3$.

The differential equation \ref{eq:adim_stress_balance} can be solved numerically with equation \ref{eq:adim_volume_conservation}.
We obtain the solution $z(t)$ by using the function \textsf{odeint} implemented in scipy \cite{Jones2001,Virtanen2020}, from which we also compute the numerical traveling duration $T_{\rm num}$.
The resulting dynamics $z(t)$ is plotted in black in figure~\ref{fig:bubble_dynamics_example}.
The fitting parameter $\xi = 32$ is used to describe all the experiments.
To circumvent the small experimental variations of the initial experimental volume, the volume for the numerical resolution is chosen to match each experiment that is precisely determined from the final $z$ position.
Figure~\ref{fig:bubble_dynamics_example} shows that the numerical solution is in good agreement with the experimental measurements.

\subsection{Approximated analytical solution}\label{sec:bubble_approx}
To get a better insight about the effect of the physical parameters, we derive an analytical solution of equation~\ref{eq:adim_stress_balance}.
To do so, we make two approximations for which the validity will be checked both against the numerical solution and the experimental results.

First, in Regime 1, we neglect the contribution of $\Omega_{\rm cap}$ in equation \ref{eq:Volume_reg1}.
This approximation is found to be lower than 10~\% on the total volume for  $R/a > 1.25$ as shown in Section 1.1 of the Supplementary Information.
In addition, we assume that the contribution of Regime 2 on the total duration of the dynamics is small.
This assumption is also expected to be realistic for an initial bubble radius $R_0$ much larger than the tube radius $a$.

Under these conditions, we have $\Omega_{\rm out}(t)  \simeq \Omega_{\rm bubble}(t) = \frac{4}{3} \pi R(t)^3 $, which simplifies the volume conservation from equation \ref{eq:adim_volume_conservation} and reads $\tilde{z} = 1 - \tilde R(\tilde t)^3 / \tilde R_0^3$.
Thus, the differential equation \ref{eq:adim_stress_balance} reduces to
\begin{equation}
    2^{3/2} \tilde{R}_0^3 = - 3 \, n^{3/2} \tilde R^{7/2} \tilde{\dot R},
\end{equation}
which leads after integration to $\tilde R^{9/2} = \tilde R_0^{9/2} - 3\sqrt{2} \,n^{-3/2}  \tilde R_0^3 \, \tilde t$, with the initial condition $\tilde R (0) = \tilde R_0$.

In this limit, the emptying duration $T_{\rm approx}$ has an analytical form that writes
\begin{equation}\label{eq:bubble_T_approx}
    T_{\rm approx} =  \frac{2\sqrt{2}}{9} \frac{\mu_\ell n^{3/2} \xi^{3/2} a}{ \gamma} \left(\frac{R_0}{a}\right)^{9/2}.
\end{equation}

Equation~\ref{eq:bubble_T_approx} indicates that the predicted timescale evolves with the number of films in the tube as $n^{3/2}$ and also  as $(R_0/a)^{9/2}$.
Therefore, we expect a significant effect of these two parameters.

 \begin{figure}
     \centering
     \includegraphics[width=.98\linewidth]{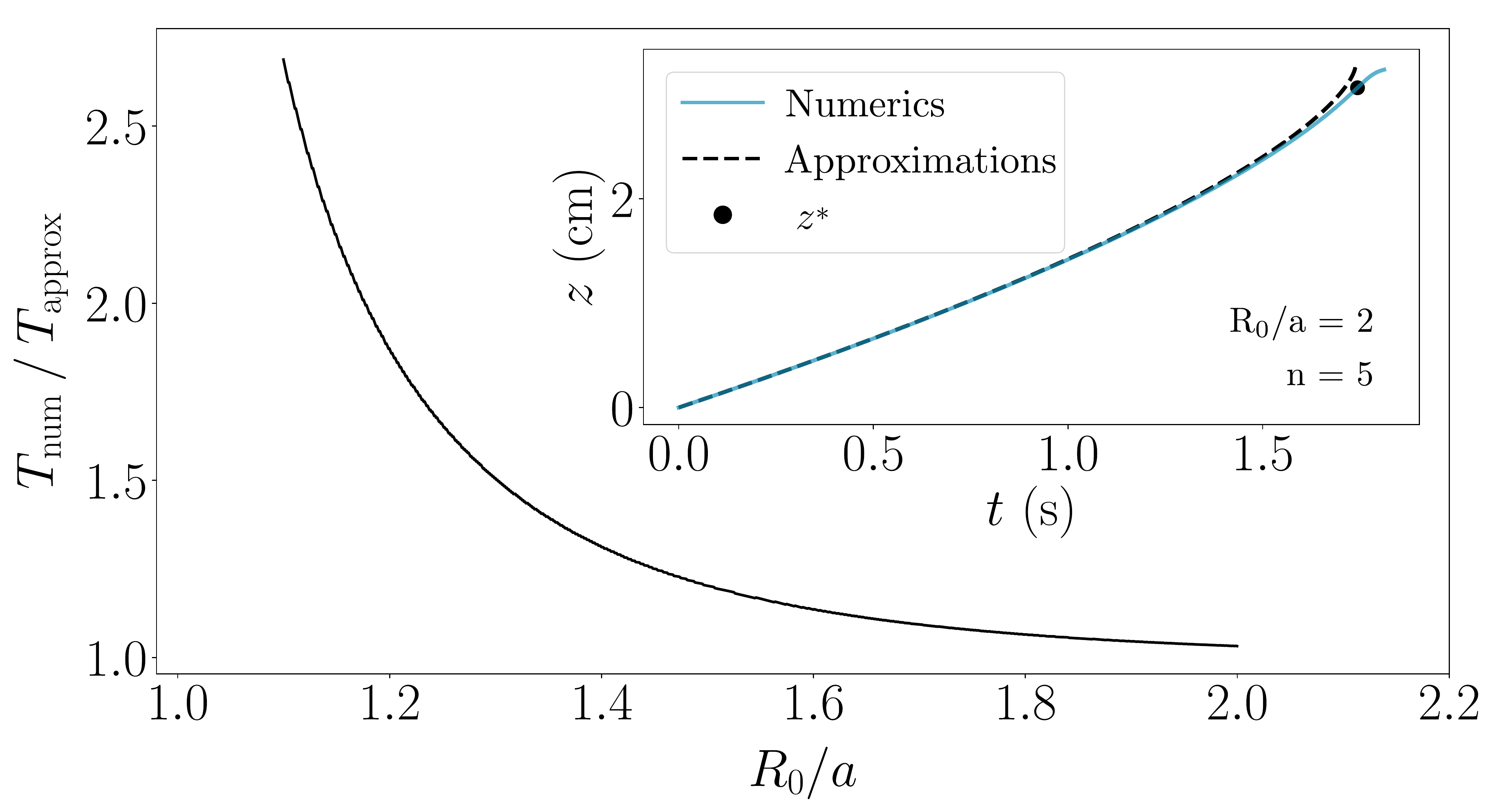}
     \caption{
     Ratio between the numerical traveling duration $T_{\rm num}$ and the approximated time $T_{\rm approx}$ as a function of the dimensionless initial bubble size $R_0/a$ for $n=5$ for $\xi = 32 $.
     We show in Section 1.2 of the Supplementary Information that this ratio is independent of $n$.
     In the inset, both numerical and approximated position of the lamella are plotted against the numerical time for $R_0/a = 2$ and $n=5$.
     The inflection point ($t^\star, z^\star$) of the curve obtained from numerics is marked with a filled circle.
     }
     \label{fig:model_precision_approx}
 \end{figure}

In figure~\ref{fig:model_precision_approx}, we analyze the impact of the approximations by comparing the numerical and the analytical durations.
The numerical solution provides a timescale that is always larger than $T_{\rm approx}$.
The typical dynamics $z(t)$ is illustrated in the inset of figure~\ref{fig:model_precision_approx}, where we observe that the approximated model stops near the inflection point $(t^\star, z^\star)$ of the numerical solution.
Thus, the approximated model predicts a shorter duration than the full resolution.
Nevertheless, we observe that the approximated duration converges rapidly toward $T_{\rm num}$ as the dimensionless radius $R_0/a$ increases.
In particular, the difference becomes less than 10~\% above $R_0/a\approx 1.7$ and is independent of the number of lamellae as shown in Section 1 of the Supplementary Information.
As a consequence, the timescale associated with the numerical resolution also scales as $n^{3/2}$.

\begin{figure}[h!]
     \centering
     \includegraphics[width=1.\linewidth]{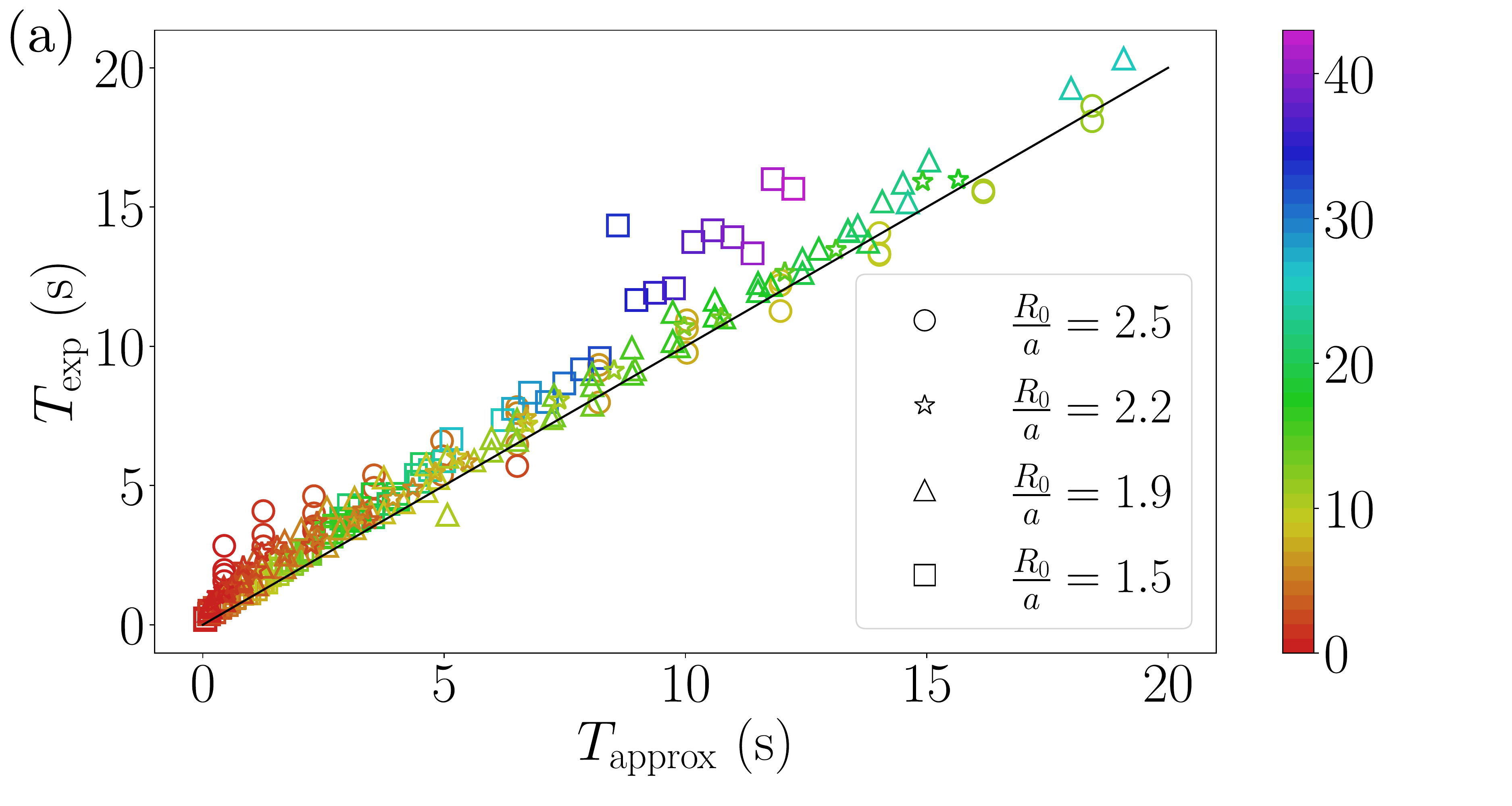}\\
     \includegraphics[width=1.\linewidth]{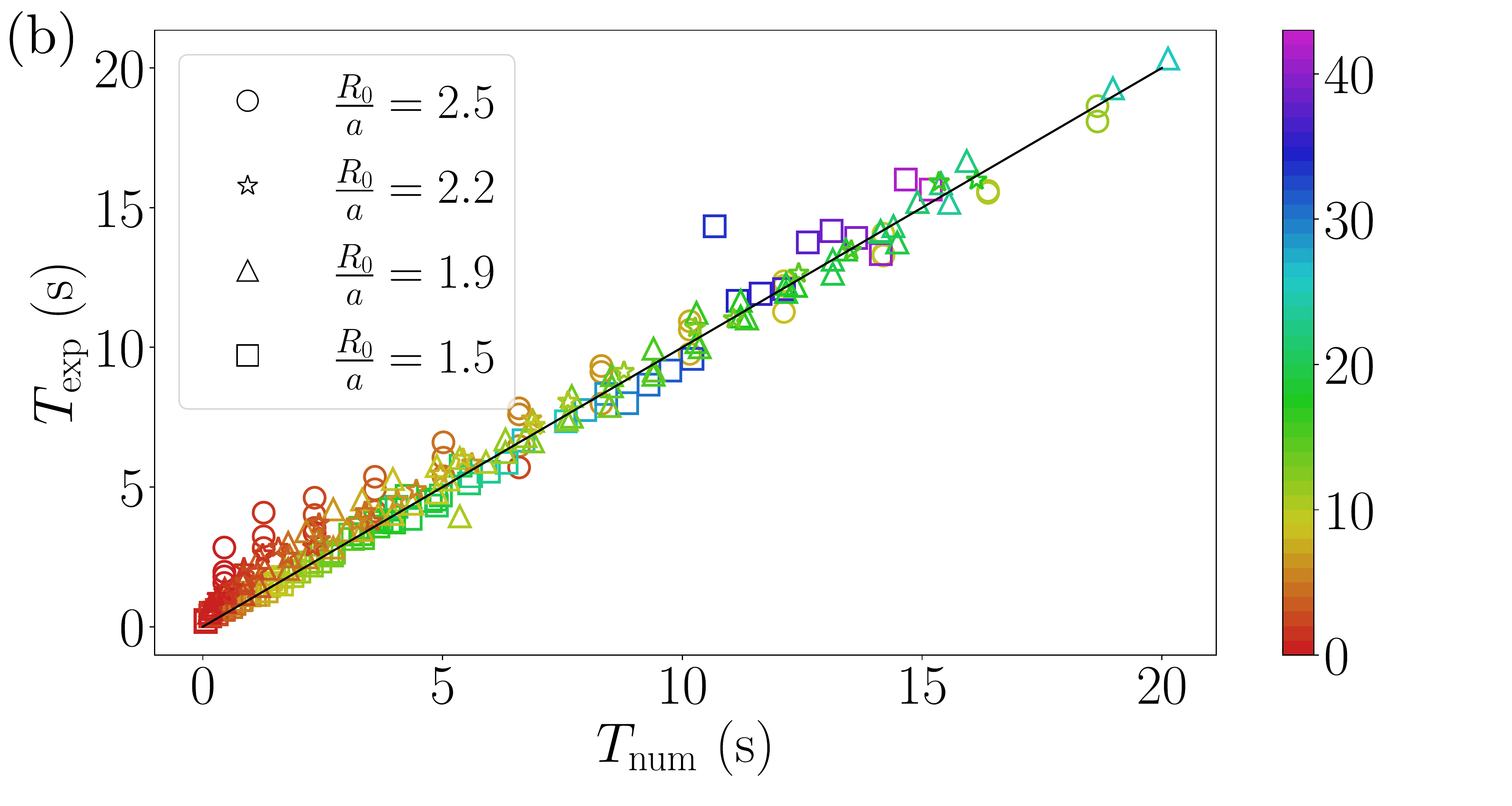}\\
     \caption{(a) Experimental emptying duration $T_{\rm exp}$ of a bubble as the function of $T_{\rm approx}$ from equation~\ref{eq:bubble_T_approx} for $\xi=32$.
     (b) Experimental emptying duration $T_{\rm exp}$ of a bubble as the function of $T_{\rm num}$ from the numerical resolution for $\xi=32$.
     On both plots, the solid black lines represent the equality between axes and each point corresponds to a bubble where the color indicates the number of lamellae in the tube from $n = 1$ to $n = 43$.
     }
     \label{fig:bubble_dynamics_duration}
\end{figure}

\subsection{Discussion}

To validate the model, we repeated the experiments described in Section~\ref{sec:exp} for bubble sizes $R_0 / a \in [1.5,1.9,2.2,2.5]$ with a total number of bubbles varying between 20 to 43.
The dynamics of the positions for all these experiments are well fitted for $\xi = 32$ (See Section 2 of the Supplementary Information).
We report in figure \ref{fig:bubble_dynamics_duration} the emptying duration $T_{\rm exp}$ of each bubble as a function of both our prediction from the approximated model and the traveling duration $T_{\rm num}$.
The equality between those times, represented by the black line, slightly underestimates most of the data especially the first bubbles (in red) and the last ones (squares in purple) for the approximated time (Fig.~\ref{fig:bubble_dynamics_duration}(a) and Section 2 of the Supplementary Information).
On the other hand, only the traveling durations $T_{\rm num}$ of the first bubbles are underestimated  (Fig.~\ref{fig:bubble_dynamics_duration}(b)).

The first bubbles have a $T_{\rm exp}$ comparable to the time needed to put the bubble in contact with the tube.
Thus, a significant part of the dynamics is not captured by the model because the initial conditions are not matching the experimental ones.
This leads to a larger $T_{\rm exp}$ than both  $T_{\rm approx}$ and $T_{\rm num}$.
Additionally, for data at $R_{\rm 0}/a$ = 1.5 (squares) there is 20~$\%$ of discrepancy between the numerical and approximated solutions compared to less than $5 \%$ for the other experiments (See figure \ref{fig:model_precision_approx}).
In figure \ref{fig:bubble_dynamics_duration}(a), the purple dots come from experiments with this aspect ratio, explaining why these measurements are more underestimated by the approximated model than the others.

In figure \ref{fig:bubble_dynamics_duration}(b), we show the comparison with the numerical prediction, which is in even better agreement for the purple points but not especially for the first ones in red, supporting  our argumentation.
Nevertheless, the analytical prediction correctly estimates the trend and the emptying timescales.

%
%
\section{Conclusion}

In this paper, we investigated the emptying of a bubble placed at the extremity of a tube.
We have shown that the bubble Laplace pressure is responsible of this motion and pushes a soap film in the tube.
This film motion dissipates energy through the liquid viscosity with a Bretherton-like behavior.
The proposed model, balancing these two forces and solved numerically, successfully describes the dynamics of this system.
Under some additional hypothesis, we obtained an analytical prediction of the emptying duration that scales with the initial bubble radius as $R_0^{9/2}$ and with the number of soap films as $n^{3/2}$.
The model is validated with a unique fitting parameter value $\xi=32$, which is a dimensionless prefactor of the viscous friction force, whose physical origin remains to be elucidated \cite{Raufaste2009,Cantat2013a,Emile2013}.

The present study focused on single bubbles deposited at the extremity of a tube.
For tubes plunged in a foam, we expect that additional parameters can affect the dynamics such as the foam rheology at play in the necessary bubble rearrangements in the foam.
Describing this dynamics will require to extend the model that we proposed, which will be the purpose of future studies.

\section*{Acknowledgments}
We thank N. Beserman and M. Dupuy for preliminary experiments, M. Hénot for useful discussions, F. Restagno for proofreading the manuscript, and L. Courbin for sharing their work prior publication.
We acknowledge for funding support from Ecole Doctorale de Physique en Ile de France (EDPIF) and from the French Agence Nationale de la Recherche in the framework of project AsperFoam - 19-CE30-0002-01.

\bibliography{biblio}

\bibliographystyle{unsrt}

\newpage\clearpage


\section*{Supplementary material}

# Supplementary information: Dynamics of bubbles spontaneously entering in a tube

Alexis Commereuc[1], Manon Marchand[1], Emmanuelle Rio[1], and François Boulogne[1]

[1]Université Paris-Saclay, CNRS, Laboratoire de Physique des Solides, 91405, Orsay, France.

## 1 Model

### 1.1 Approximation of a negligible $\Omega_{\text{cap}}$

The first of the two approximations made for deriving the approximated shrinking duration considers a negligible contribution of the cap volume compared to the total volume expressed as

$$\Omega_{\text{out}}^{\text{Regime1}}(t) = \frac{4}{3}\pi R(t)^3 - \Omega_{\text{cap}}(t), \tag{S1}$$

where

$$\Omega_{\text{cap}}(t) = \frac{\pi R(t)^3}{3}(1-\mathcal{G})^2(2+\mathcal{G}) \tag{S2}$$

with $\mathcal{G}(a, R(t)) = \sqrt{1 - a^2/R(t)^2}$.

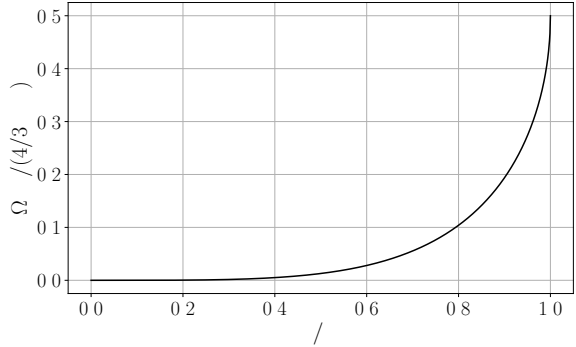

Figure S1: Ratio of the cap volume to the total volume $\Omega_{\text{cap}}/(4/3\pi R^3)$ as a function of the inverse of the dimensionless bubble radius $a/R$.

In figure S1, we plot the ratio between the two terms on the left hand side of equation S1, *i.e.* $\Omega_{\text{cap}}/(4/3\pi R^3)$ as a function of $a/R$. We observe that under the approximation, the volume $\Omega_{\text{out}}^{\text{Regime1}}(t)$ is overestimated by less than 10 % for $a/R < 0.8$ *i.e.* $R/a > 1.25$.

### 1.2 Effect of the number of lamellae $n$ on the ratio $T_{\text{num}}/T_{\text{approx}}$

In figure S2 (a) we first plot both the numerical and approximated times presented in the model as the function of $n^{3/2}$. We notice that the approximated duration is smaller than the numerical prediction, and the difference is increasing with $n^{3/2}$.

Because both duration scale as $n^{3/2}$, we can expect a constant ratio between them. This is evidenced in figure S2 (b) for different dimensionless aspect ratios $\tilde{R}$. As presented in fig 5, this ratio depends on the aspect ratio but is constant with the number of lamellae explaining why the approximated time for large $n$ values, which exists only for small aspect ratio, is not in good agreement with the experiment in fig 6.

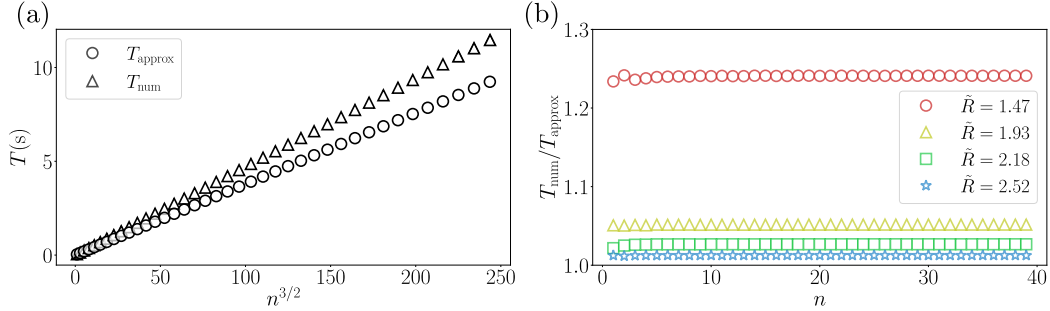

Figure S2: (a) Both numerical and approximated times needed for one bubble to empty in the tube for $\tilde{R} = 1.74$ as a function of the number of lamellae $n$. (b) Ratio between those two times for different aspect ratio $\tilde{R}$ as a function of $n$.

## 2 Experimental results

### 2.1 Reproducibility of the experiment

We perform all our data set three times and we illustrate the reproducibility of our experiment in figure S3 where we plot the results for three different experiments with $\Omega_{\text{bubble}} = 0.42 \pm 0.03$ mL and four different $n$ value. We can see that each experiment is well reproducible. In the next figure, we show only one experimental data set for legibility.

### 2.2 Individual bubble dynamics

To show the agreement of the numerical model, we present the individual fitted dynamics of the bubble shrinkage for three different aspect ratio: $\tilde{R} = 1.47$ in figure S4, $\tilde{R} = 1.93$ in figure S5, $\tilde{R} = 2.18$ in figure S6. In each case, data for the different number $n$ of lamellae are presented. We notice the disagreement for the first bubbles (typically $n < 3$) due to the time needed to put the bubble in contact with the tube that is comparable to the shrinkage duration.

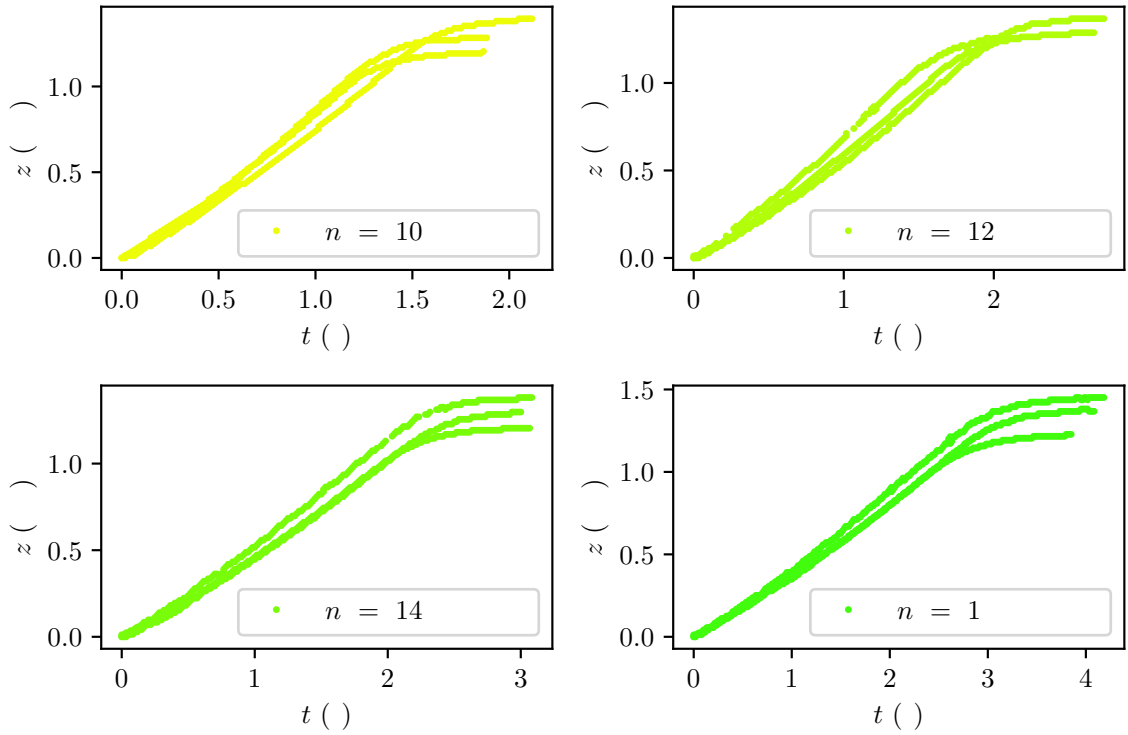

Figure S3: Dynamics of the lamella for 3 bubbles of volume $\Omega_{\text{bubble}} = 0.42 \pm 0.03$ mL. The dots are experimental data with the same color code as in figure 6(b). Each plot has a unique value of lamellae $n$.

3

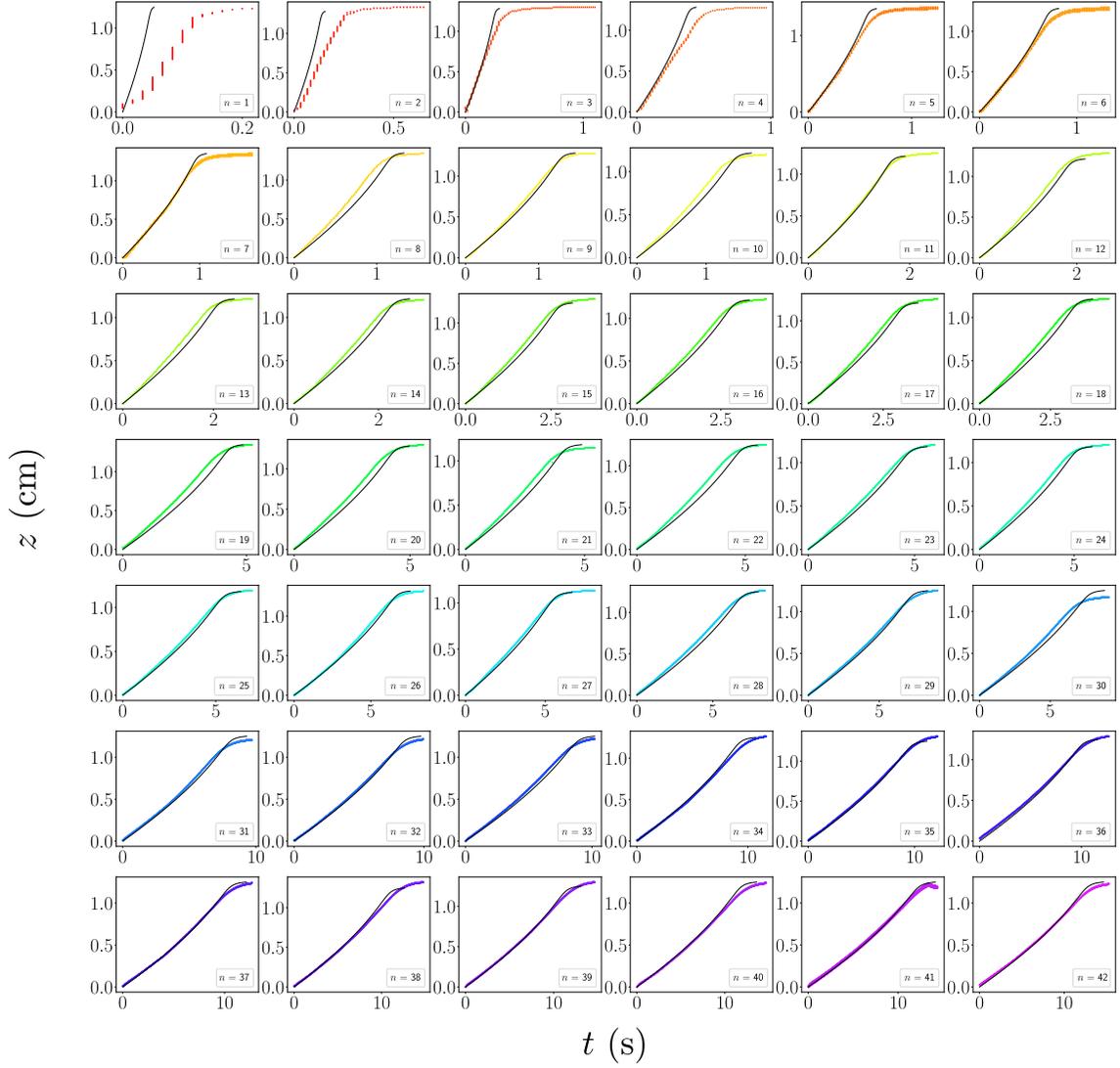

Figure S4: Dynamics of the lamella for 42 bubbles of volume $\tilde{R} = 1.47$. The colored dots are experimental data with the same correspondence as in figure 6(b) and the black lines are the numerical model.

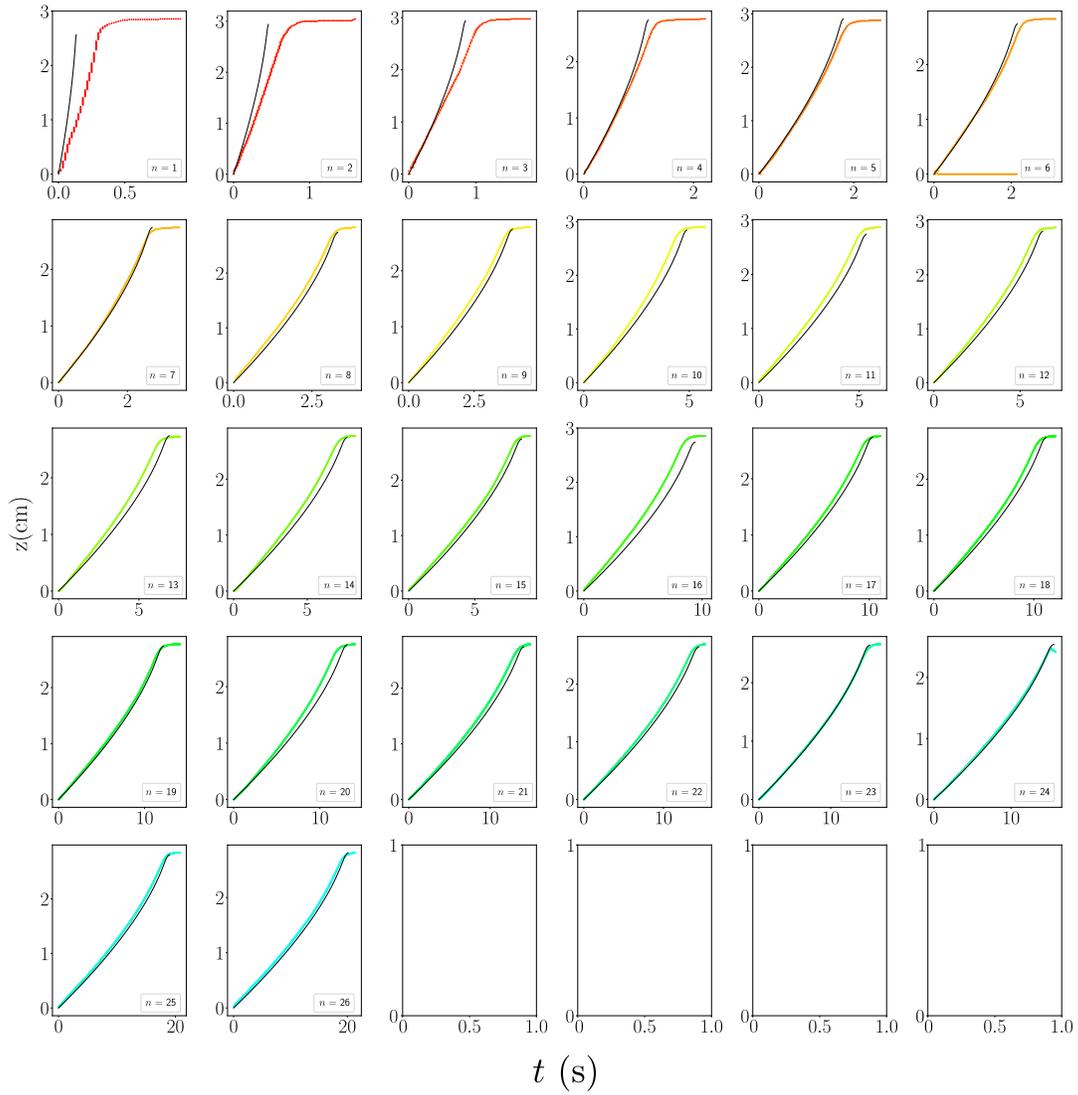

Figure S5: Dynamics of the Incompressibilitylamella for 26 bubbles of volume $\tilde{R} = 1.93$. The colored dots are experimental data with the same correspondence as in figure 6(b) and the black lines are the numerical model.

5

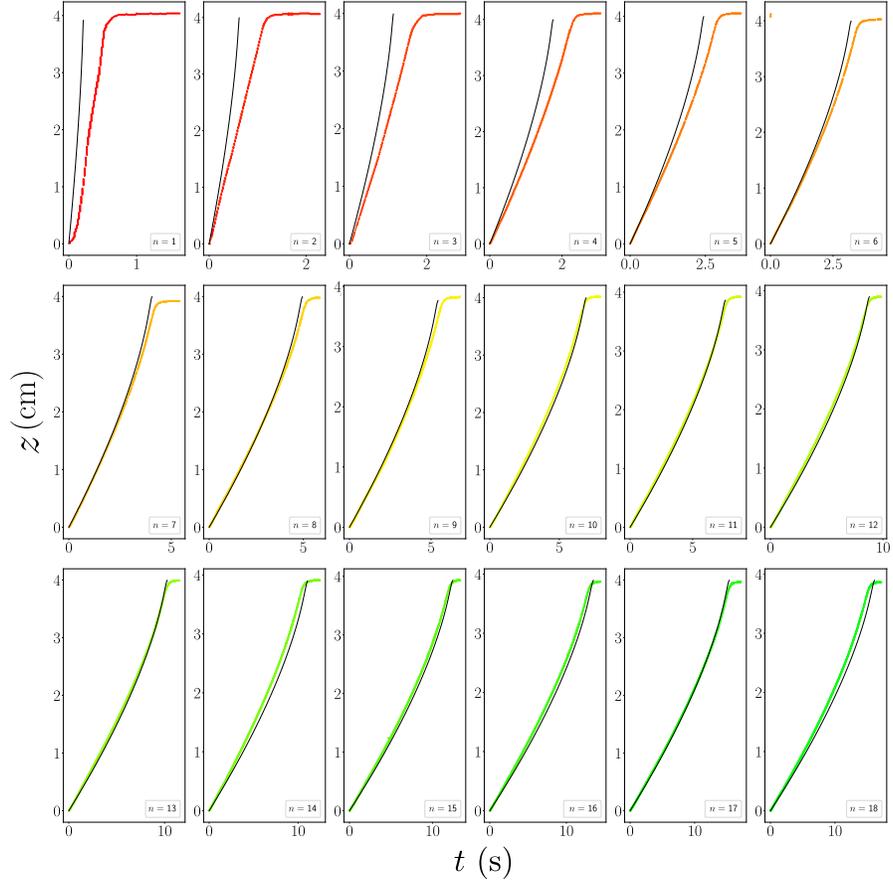

Figure S6: Dynamics of the lamella for 18 bubbles of volume $\tilde{R} = 2.18$. The colored dots are experimental data with the same correspondence as in figure 6(b) and the black lines are the numerical model.

6



\end{document}